\documentclass{amsart}
\usepackage{amsmath}
\usepackage{amssymb}
\usepackage{bm}
\usepackage{graphicx}

\begin{document}

\title[A mathematical walk into the paradox of Bloch oscillations]{A mathematical walk into the paradox \\ of Bloch oscillations }

\author{Luigi Barletti}

\address{Dipartimento di Matematica e Informatica ``Ulisse Dini'', 
Universit\`a degli Studi di Firenze}


\maketitle

\begin{abstract}
We describe mathematically the apparently paradoxical phenomenon that an electronic current in a semiconductor can flow {\em because} of
collisions, and not despite them.
A transport model of charge transport in a one-dimensional semiconductor crystal is considered, where each electron follows the periodic hamiltonian trajectories, 
determined by the semiconductor band structure, and undergoes non-elastic collisions with a phonon bath. 
Starting from the detailed phase-space model, a closed system of ODEs is obtained for averaged quantities.  
Such a simplified model is nevertheless capable of describing transient Bloch oscillations, their damping and the consequent onset of a steady current flow, 
which is in good agreement with the available experimental data. 
\end{abstract}

\section{Introduction}
\label{S1}
Microelectronics, whose fundamental importance for the modern civilization needs not to be stressed, is based on electronic currents flowing 
in semiconductor materials. 
Such currents are the macroscopic manifestation of the microscopic dynamics of electrons inside the semiconductor and of their interaction with 
external perturbations, such as electrostatic fields, light, heat, mechanical pressure, etc.. 
Understanding the behaviour of electrons in semiconductors has become therefore one of the most important branches of physics, which is known 
as {\em solid-state physics}.
This is very well known. 
But what is probably less known is the curious phenomenon underlying all of this: the electronic current 
is made possible by the same collisions that hinder it.
The aim of these notes is to illustrate this apparent paradox by means of a simple mathematical model of electron transport in a semiconductor crystal
under the action of an external electrostatic field. 
\par
\smallskip
A semiconductor is a crystalline solid made by ions periodically arranged in space and held together by covalent bonds.
So, what a ``free'' (non-bond) electron feels inside a semiconductor crystal is a periodic electrostatic potential generated by the crystal ions.
Now, according to the basic laws of quantum mechanics, the possible energies $E$ of a particle with potential energy $W$ are given by the
stationary Schr\"odinger equation
$$
   -\frac{\hbar^2}{2m} \Delta \psi + W\psi = E\psi
$$
which is the eigenvalue equation for the quantum hamiltonian $-\frac{\hbar^2}{2m} \Delta  + W$. 
For example, in the case of a free particle ($W = 0$) one finds a continuum of possible energies (namely, $E\geq 0$), while in the case of a harmonic trap
($W$ is a harmonic potential) one finds a sequence of discrete levels.
When $W$ is a periodic potential, which is the case of our electron in the crystal lattice, one finds that the possible energies  form a sequence of intervals,
called {\em energy bands}, alternated with {\em forbidden bands}.
Therefore, the periodic case is, so to say, a mix of the free and the trapped  particle cases. 
Moreover, and this point will be central for all that follows, each energy band is a periodic function $\mathcal{E}_n(p)$ of the electron momentum $p$,
where $n$ is an integer labelling the bands.
\par
What we have said so far is not much different from what could be said for metals. 
What distinguishes a semiconductor from a metal is that a semiconductor possesses a {\em energy gap}, that is a forbidden band which is placed
more or less at the level of the Fermi energy.
This means (by simplifying a little) that the energy bands below the gap are statistically ``fully occupied'' by electrons and cannot contribute to the 
conduction of current \cite{AM1976}. 
On the contrary, the energy band just above the Fermi level is only partially occupied and its electrons can produce a current 
(not for nothing it is called {\em conduction band}).
The fact that in a semiconductor the conduction band is energetically isolated will allow us to consider, in first approximation, this single energy band 
in the mathematical model that will be discusse hereafter.
This is why our model applies to semiconductors and not to metals, even though many of the considerations that will be made are also true for metals.
\par 
Let us then consider an electron in the conduction band and assume that a constant electrostatic force is exerted to it (e.g., by an applied voltage).
Such electron will be uniformly accelerated, or, more precisely, its momentum $p$ will increase (or decrease, according to the direction of the force) linearly with time.
But, because of the periodicity of the energy band as a function of $p$, the conservation of energy implies that also the potential energy due to the 
external field must vary periodically, which means that the electron position will vary periodically, so that electron will start moving back and forth.
Such periodic motion is called {\em Bloch oscillations}, hereafter abbreviated with BO. 
Since it is extremely difficult to observe the BO in bulk semiconductors (the Fourier reciprocity between space and momentum periods makes the latter
too large to be entirely spanned by electrons, in normal conditions), they have been observed in artificial semiconductor ``superlattices'' \cite{DOK1995,Leo92}.
We will come back on this point in Section \ref{S5}.
\par
\begin{figure}
\begin{center}
\includegraphics[scale=0.9]{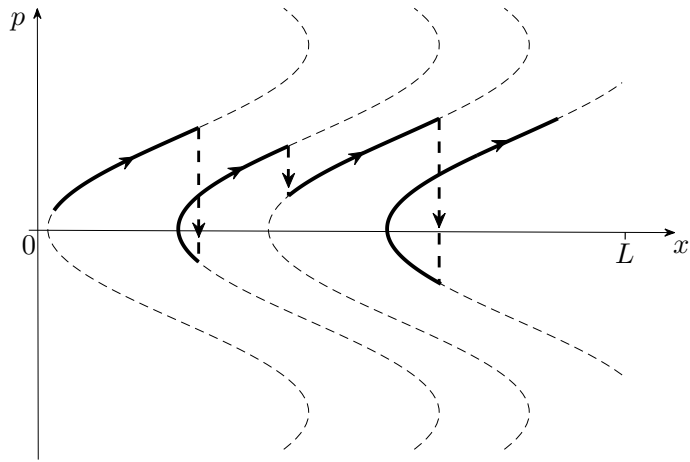}
\caption{A pictorial representation of the BO paradox: an electron, which would be otherwise trapped on a hamiltonian periodic trajectory in phase-space (dashed curves), 
jumps from one trajectory to another thanks to inelastic collisions (dashed arrows). 
This produces an average advance towards the direction of the field (here directed rightward).}
\label{fig1}
\end{center}
\end{figure}

Then, the purely conservative hamiltonian dynamics of an electron in a semiconductor is an oscillatory motion which would prevent it from originating
any current.
So, how is it possible that an electric current can actually flow in a semiconductor?
The answer is in the {\em obstacles} that the electron finds on his path.  
Such obstacles, the interactions with whom will be generically called ``collisions'',  can be of various kind, the most important being the {\em phonons}, 
i.e.\ the thermal vibrations of the crystal lattice.
A collision with a phonon is an inelastic process which makes the electron (instantaneously and locally, in first approximation) change momentum and loose energy,
thus jumping on a different hamiltonian trajectory. 
If the collision are frequent enough (but not too much), they will allow the electron to jump from one trajectory to another, which is what really makes it 
advance in the direction of the field, see Figure \ref{fig1}.
Hence, we come to the paradoxical conclusion that the electronic current can flow {\em because} of collisions, and not despite them.
\par
\smallskip
The ideas that we have just sketched, will be made more precise in the rest of the paper with the introduction of a transport model that includes both the BO dynamics 
and the collisions.
In Section \ref{S2} we construct the transport model under the form of a semiclassical Boltzmann equation for the electronic density in phase-space. 
After non-dimensionalisation, in Section \ref{S3} a system of fluid-dynamic, Euler-like, equations are derived from the transport equation.
The closure of such system is then obtained in the limit when the number of collision, with respect to the time that the electron momentum takes to span the BO period, 
goes to infinity.
This leads to a drift-diffusion equation and, therefore, to the existence of a current.
In Section \ref{S4}, we take space-averages of the fluid variables and find a system of ODEs which has the form of a forced and damped harmonic oscillator.
The dynamics described by this system consists in an oscillatory transient, dominated by BO, the oscillation damping and the consequent onset of a steady current flow.
The obtained solution is then discussed in relation with the available experimental data in the last section, Section \ref{S5}.
\section{Transport model for the electron dynamics}
\label{S2}
Let us consider a simple model of a semiconductor by assuming that the crystal lattice has only one spatial dimension ad that the charge transport takes place in
a single conduction band $\mathcal{E}(p)$. 
Moreover, let us assume that the energy of the electrons in this band takes the form: 
\begin{equation}
\label{band}
\mathcal{E}(p) = E_0\left(1-\cos\left(\frac{p}{p_0}\right)\right),
\end{equation}
which is known as {\em Kronig-Penney} dispersion relation \cite{BJ2000}. 
This is equivalent to assuming that the lattice periodic potential has the shape of a ``square wave''.
This is a simplification in general but, in the case of a semiconductor superlattice (see Section \ref{S5}), 
it can be very close to the real situation.
We remark that this energy is a periodic function of the electron momentum $p$, with a period of $2\pi p_0$ and an amplitude of $E_0$. 
In a semiclassical perspective, the electron velocity as a function of $p$ is given by 
\begin{equation}
\label{vel}
\frac{\partial\mathcal{E}}{\partial p} =  \frac{E_0}{p_0} \sin\left(\frac{p}{p_0}\right).
\end{equation}
and the electron effective mass \cite{CMP11} is given by
\begin{equation}
\label{em}
\frac{1}{m^*} = \frac{\partial^2\mathcal{E}}{\partial p^2} \Big\vert_{p = 0} =  \frac{E_0}{p_0^2}.
\end{equation}
Note that the effective mass is, indeed, a quantity that plays the role of the mass in the quadratic approximation of the energy:
$$
\mathcal{E}(p) \approx \frac{p^2}{2m^*}
$$
(physically speaking, for low energies, a particle in a periodic potential behaves as a free particle but with a different mass).
\par
On the other hand, we assume that the electron is also subject to an external electrostatic field with a linear profile 
of the form
\begin{equation}
\mathcal{U}(x) = -Fx.
\end{equation}
This of course corresponds to a constant force $F$ exerted by an applied voltage $V$ such that $qV = -FL$, 
where $q$ is the elementary charge and $L$ is the device length.
\par
In this picture, the electron (classical)  Hamiltonian can be written as
\begin{equation}
\mathcal{H}(x,p) =  \mathcal{E}(p)  + \mathcal{U}(x) 
= E_0\left(1-\cos\left(\frac{p}{p_0}\right)\right) - Fx,
\end{equation}
where $p$ is a periodic variable, of period $2\pi p_0$, and $x \in [0,L]$.
The corresponding Hamilton equations are
\begin{equation}
\left\{
\begin{aligned}
&\dot{x} = \frac{\partial\mathcal{H}}{\partial p}  = \frac{E_0}{p_0} \sin\left(\frac{p}{p_0}\right),
\\[6pt]
 & \dot{p} = -\frac{\partial\mathcal{H}}{\partial x} = F,
\end{aligned}
\right.
\end{equation}
describing the unperturbed BO motion of the electron.
\subsection{Transport equation}
\label{S2.1}
Let $f(x,p,t)$, $x\in [0,L]$, $p\in [-\pi p_0, +\pi p_0]$, $t \geq 0$, be the electron phase space density function at time $t$. 
As the crystal has only one spatial dimension, the spatial electron density $\rho$ can be expressed as
\begin{equation}
\label{rhodef}
    	\rho(x,t) = \int_{-\pi p_0}^{\pi p_0}f(x,p,t)\,dp.
\end{equation}
and the electronic current density can be expressed as
\begin{equation}
\label{jdef}
    	j(x,t) = \frac{E_0}{p_0} \int_{-\pi p_0}^{\pi p_0} \sin\left(\frac{p}{p_0}\right)\,f(x,p,t)\,dp.
\end{equation}
The non-collisional, semiclassical transport equation for $f$ is the Liouville equation for the Hamiltonian $\mathcal{H}$, namely,
\begin{equation}
\label{Liouville}
    	 \frac{\partial f}{\partial t} + \dot{x}\frac{\partial f}{\partial x} + \dot{p}\frac{\partial f}{\partial p} 
	 =  \frac{\partial f}{\partial t} + \frac{E_0}{p_0} \sin\left(\frac{p}{p_0}\right)\frac{\partial f}{\partial x} + F\frac{\partial f}{\partial p} = 0.
\end{equation}
This simply expresses the fact that the phase-space distribution is constant along the newtonian trajectories of the system. 
In particular, the distribution remains constant on the phase-space curves of constant energy $E$:
\begin{equation}
\label{orbits}
    	E_0\left(1-\cos\left(\frac{p}{p_0}\right)\right) - Fx = E.
\end{equation}
Equation \eqref{Liouville} describes a purely hamiltonian evolution of the electron population. 
In semiconductors, however, electrons undergo non-elastic interactions with the lattice vibrations (that can be described in terms of quasi-particles known
as {\em phonons}).
A complete description of the interactions between electrons and phonons would require a detailed scattering operator that includes the different branches 
of the phonon dispersion relations. 
Here, for the sake of simplicity, we shall assume that such interactions can be described by means of a space-homogeneous, linearized Boltzmann collisional 
operator of the following form:
\begin{equation}
\label{Qdef}
 \frac{1}{\tau_0} \int_{-\pi p_0}^{\pi p_0} b_0(p,p')\left[ \mathcal{M}(p) f(x,p',t) - \mathcal{M}(p') f(x,p,t) \right] dp',
\end{equation}
where $\tau_0$ is the typical collision time (or, in other words, $1/\tau_0$ is the collision frequency),  $b_0(p,p') = b_0(p',p) $ is the scattering kernel
(i.e., the probability of the electron momentum $p'$ to be scattered to a new momentum $p$, following a collision) and
\begin{equation}
\label{cMdef}
 \mathcal{M}(p)  =  \frac{1}{\sqrt{2\pi m^* k_BT}}\, e^{-\frac{p^2}{2m^*k_BT}}
\end{equation}
is the Maxwellian distribution.
In the last expression, $m^*$ is the effective electron mass \eqref{em}, $k_B$ is the Boltzmann constant and $T$ is the temperature of the phonon bath.  
Note that the collisions term described by \eqref{Qdef} can be interpreted as follows: electrons collide with a typical frequency $1/\tau_0$ and
the effect of collisions is  to re-distribute the electron momenta according to a Maxwellian (thermal) distribution at the temperature of the phonon bath.
\par
The complete transport model is now obtained by adding the collision term \eqref{Qdef} to the Liouville equation \eqref{Liouville}, 
which results in the transport equation
\begin{equation}
\label{Boltzmann}
\frac{\partial f}{\partial t} + \frac{E_0}{p_0} \sin\left(\frac{p}{p_0}\right)\frac{\partial f}{\partial x} + F\frac{\partial f}{\partial p} 
=   \frac{1}{\tau_0} \int_{-\pi p_0}^{\pi p_0} b_0 (p,p')\left[ \mathcal{M}(p) f(p') - \mathcal{M}(p') f(p) \right] dp',
\end{equation}
(in the integral, the space and time variable have been omitted).
This equation is a variant of what is sometimes called ``semiclassical Boltzmann equation'' \cite{AM1976}.
\par
An important remark is necessary here.
The choice of a Maxwellian as the thermal distribution may seem in contradiction with the periodicity of the variable $p$. 
Actually, the correct equilibrium distribution would be 
$$
   c\,e^{-\frac{1}{k_BT}\mathcal{E}(p)} ,
$$
where $c$ is the suitable normalization constant and $\mathcal{E}(p)$ is given by \eqref{band}, which is indeed periodic.
However, in standard situations, the thermal electron momentum $p_\mathrm{th} = \sqrt{m^*k_BT}$ is much smaller than $p_0$, 
which means that it is reasonable to assume 
\begin{equation}
\label{Mcond}
 \frac{p_\mathrm{th}}{p_0} = \frac{\sqrt{m^*k_BT}}{p_0} = \sqrt{\frac{k_BT}{E_0}} \ll 1
\end{equation}
(recalling the definition \eqref{em} of the effective mass $m^*$).
Under such condition, $\mathcal{E}(p)$ is very close to its quadratic approximation $\frac{p^2}{2m^*}$, which leads to the Maxwellian \eqref{cMdef}.
This also means that the Maxwellian is narrow enough that it its practically zero at the period endpoints $\pm\pi p_0$ and, therefore, it can be reasonably 
considered as periodic.
\subsection{The evolution problem}
\label{S2.2}
For its physical and mathematical consistency,  equation \eqref{Boltzmann} must be supplemented with suitable conditions on the inflow part of the boundary:
\begin{equation}
\label{BC}
\left\{
\begin{aligned}
&f(0,p,t) = \phi_l(p), \quad \text{if $p>0$,}
\\[6pt]
&f(L,p,t) = \phi_r(p), \quad \text{if $p<0$,}
\end{aligned}
\right.
\end{equation}
where $\phi_l(p)$ and $\phi_r(p)$ are the assigned inflows of electrons from the left and from the right, respectively.
Typically, they are chosen as Maxwellians distributions, at the temperature $T$, with densities corresponding to the electronic densities of the 
metallic contacts \cite{Frensley87}.
\par
Moreover, since $p$ is a periodic coordinate, we have also to impose the periodicity condition on the momentum boundary:
\begin{equation}
\label{PC}
  f(x,\pi p_0,t) =  f(x,-\pi p_0,t).
\end{equation}
Finally, we have to assign the density function at the initial time (say, $t = 0$):
\begin{equation}
\label{IC}
  f(x,p,0) =  f_{\mathrm{in}}(x,p).
\end{equation}
It can be proven that, under suitable regularity assumptions on the data,  
the initial-boundary value problem \eqref{Boltzmann}-\eqref{BC}-\eqref{PC}-\eqref{IC} is well-posed as an evolution problem in the Banach space 
$L^1(\mathbb{R}_x\times\mathbb{R}_p)$ of integrable functions on phase-space \cite{LB2000,Belleni79,TilliTesi}.
Such evolution problem can be formally written as 
\begin{equation}
\left\{
\begin{aligned}
 &\dot f(t) = Af(t) + Bf(t), \\ &f(0) = f_{\mathrm{in}}
\end{aligned}
\right.
\end{equation}
where  
$$
  Af  =  \frac{E_0}{p_0} \sin\left(\frac{p}{p_0}\right)\frac{\partial f}{\partial x} + F\frac{\partial }{\partial p}
$$
is the Liouville operator (defined on a suitable domain that contains the boundary conditions \eqref{BC}-\eqref{PC}) and
$$
 Bf =  \frac{1}{\tau_0} \int_{-\pi p_0}^{\pi p_0} b_0 (p,p')\left[ \mathcal{M}(p) f(p') - \mathcal{M}(p') f(p) \right] dp'
$$
is the collisional operator. 
Under reasonable assumptions on the scattering kernel $b_0 (p,p')$, $B$ is a bounded perturbation and the solution to the evolution
problem can be represented as a Dyson-Phillips series (here written in the simpler case of homogeneous boundary conditions $\phi_l= \phi_r = 0$):
\begin{equation} 
\label{DyPhi}
\begin{aligned}
 f(t) &= T(t)\, f_{\mathrm{in}}
\\
        &+ \int_0^t T(t-s_1)\, B\, T(s_1)\, f_{\mathrm{in}}\,ds_1
\\
        &+ \int_0^t \int_0^{s_1} T(t-s_1)\, B\, T(s_1-s_2) B\, T(s_2)\, f_{\mathrm{in}}\,ds_1ds_2
\\[4pt]
        &+ \cdots.
\end{aligned}
\end{equation}
Here, $T(t)$ is the evolution semigroup generated by $A$, representing therefore the non-collisional dynamics (pure streaming along the newtonian trajectories), 
and each term of the series  corresponds to the contributions of electrons that have undergone 0, 1, 2, \ldots collisions in the time interval $[0,t]$.
Equation \eqref{DyPhi} is to be considered as the rigorous mathematical representation of the dynamics pictured in Figure \ref{fig1}.
\subsection{Non-dimensionalisation}
\label{S2.3}
We shall now write eq.\ \eqref {Boltzmann} in non-dimensional form. 
Let $x_0$, $p_0$, $t_0$ and $f_0$ by reference length, momentum, time and density, respectively (the reference momentum is the same $p_0$ appearing in
the energy band \eqref{band}).
The corresponding non-dimensional variables are given by
$$
   \hat x = x/x_0, \qquad \hat p = p/p_0,  \qquad \hat t = t/t_0 ,
$$
and the non-dimensional distribution function is given by
$$
  \hat f (\hat x, \hat p, \hat t) = f(x_0\hat x, p_0\hat p, t_0 \hat t ) /f_0.
$$
Making these substitutions into \eqref{Boltzmann} and multiplying by $t_{0}$ yields
\begin{multline}
\label{aux}
f_0 \frac{\partial \hat f}{\partial \hat t} + \frac{t_0 f_0 E_{0} }{p_{0} x_{0}} \sin (\hat p)  \frac{\partial \hat f}{\partial \hat x} 
+ \frac{f_0 t_{0} F}{p_{0}} \frac{\partial \hat f}{\partial \hat p} 
\\
=   \frac{t_0 f_0 p_0}{\tau_0} \int_{-\pi}^\pi  b_0(p_0 \hat p,p_0 \hat p')\left[ \mathcal{M}(p_0p) \hat f(\hat p') - \mathcal{M}(p_0 \hat p') \hat f(\hat p) \right] d\hat p',
\end{multline}
where we also performed the change of integration variable $\hat p' = p'/p_0$. 
Now, it is natural to chose as reference length $x_{0} = L$, the device length, while it is clear than the choice of $f_0$ is irrelevant, 
as it multiplies every term because of the linearity of the equation.
Moreover, as reference time we choose 
$$
  t_{0} = \frac{p_{0}}{F}.
$$
This can be interpreted as follows: since the electron is uniformly accelerated by the constant force $F$, and then $F$ is the proportionality constant
between momentum variations and time variations, $2\pi t_0$ represents the time it takes to the electron momentum to span the energy-band period 
$2\pi p_0$.
Let us now introduce  the non-dimensional scattering kernel 
$$
  b(\hat p, \hat p') = b_0(p_0\hat p,p_0 \hat p') 
$$
and the non-dimensional Maxwellian
\begin{equation}
\label{Mdef}
M(\hat p) = p_0\,\mathcal{M}(p_0 \hat p) =  \frac{p_0}{\sqrt{2\pi m^* k_BT}}\, e^{-\frac{p_0^2 \hat p^2}{2m^*k_BT}} 
=  \frac{1}{\sqrt{2\pi\sigma^2}}\, e^{-\frac{\hat p^2}{2\sigma^2}} ,
\end{equation}
where, obviously, 
$$
  \sigma^2 = \frac{m^*k_BT}{p_0^2} = \frac{k_B T}{E_0}.
$$
(note that the condition \eqref{Mcond} translates into $\sigma \ll 1$).
Then, equation \eqref{aux} can be rewritten as follows
\begin{equation}
\label{BE}
\frac{\partial f}{\partial t} + \alpha \sin (p)  \frac{\partial f}{\partial x}  +  \frac{\partial f}{\partial p} 
=   \frac{1}{\tau} \int_{-\pi}^\pi b(p,p')\left[ M(p) f(p') - M(p') f(p) \right] dp',
\end{equation}
where, for the sake of clearness, we have dropped the hats everywhere (so that the new, non-dimensional variables are now indicated 
by the same symbols as the dimensional one) and we have defined the non-dimensional parameters
\begin{equation}
\alpha =  \frac{t_0E_{0}}{p_0 L} = \frac{E_{0}}{F L},  \qquad \tau = \frac{\tau_0}{t_0} = \frac{\tau_0 F}{p_0}.
\end{equation}
These parameters are, respectively, the ratio between the width of the energy band and the energy of the external field, and the scaled collision time,
namely, the ratio between the typical collision time and the band-spanning time.
\section{Fluid equations}
\label{S3}
In order to arrive at a macroscopic description, i.e.\ a fluid-dynamical description based on local averages, 
such as density and current, we shall take moments of the non-dimensional transport equation  \eqref{BE} with respect to the 
momentum variable $p$. 
\subsection{An Euler-like system}
\label{S3.1}
First of all, we calculate the integral with respect to $p \in [-\pi,\pi]$ of both sides of the transport equation \eqref{BE}.
In particular, note that at the right-hand side we obtain 
\begin{equation}
\int_{-\pi}^\pi \int_{-\pi}^\pi b(p,p')\left[ M(p) f(p') - M(p') f(p) \right] dp'\, dp = 0,
\end{equation}
owing to $b(p,p') = b(p',p)$.
This reflects the fact that collisions locally conserve the number of particles. 
Then, the $p$-average of \eqref{BE} yields the continuity equation
\begin{equation}
\label{ME1}
\frac{\partial \rho }{\partial t} + \frac{\partial j}{\partial x} = 0,
\end{equation}
where 
\begin{equation}
\label{rhoj}
    	\rho(x,t) = \int_{-\pi}^{\pi}f(x,p,t)\,dp,
	\qquad
	 j(x,t) = \int_{-\pi }^{\pi } \alpha\sin(p)\,f(x,p,t)\,dp
\end{equation}
are the non-dimensional counterparts of \eqref{rhodef} and \eqref{jdef}.
We remark that eq.\ \eqref{ME1} has been obtained by assuming the periodicity of $f$ as a function of $p$ 
(making the boundary term vanish in the integration by parts).
\par
To obtain an equation for $j$, we iterate the procedure by multiplying both sides of eq.\ \eqref{BE} by $\alpha\sin(p)$ and taking the $p$-average.
In particular, by assuming $b$ to be an even function,
\begin{equation}
  b(p,p') = b(-p,p'),
\end{equation}
at the right-hand side we obtain 
\begin{multline*}
\int_{-\pi}^\pi \frac{ \alpha}{\tau} \sin(p) \int_{-\pi}^\pi\, b(p,p')\left[ M(p) f(p') - M(p') f(p) \right] dp' \, dp 
\\
= - \frac{1}{\tau} \int_{-\pi}^\pi   \alpha \gamma(p)\,\sin(p)\,f(p)\,dp,
\end{multline*}
where 
\begin{equation}
  \gamma(p) = \int_{-\pi}^\pi   b(p,p')\, M(p')\,dp'.
\end{equation}
Then:
\begin{equation}
\label{ME2}
\frac{\partial j }{\partial t}  +   \frac{\partial}{\partial x} \int_{-\pi}^{\pi} \alpha^2 \sin^2(p)\,f(x,p,t)\,dp - \kappa(x,t) = 
- \frac{1}{\tau} \int_{-\pi}^\pi \alpha  \gamma(p)\,\sin(p)\,f(p)\,dp,
\end{equation}
where we introduced the new macroscopic quantity
\begin{equation}
\label{kdef}
 \kappa(x,t) =  \int ^{\pi} _{-\pi} \alpha \cos (p)\, f(x,p,t)\, dp.
\end{equation}
Recalling \eqref{band}, we see that $\kappa$ is closely related to a local energy density.
\par
In order to obtain an Euler-like system for the macroscopic quantities $\rho$, $j$ and $\kappa$, we iterate once again the procedure by taking the $p$-average
of eq.\ \eqref{BE} multiplied by $\alpha\cos(p)$.
This yields
\begin{equation}
\label{ME3}
\frac{\partial k}{\partial t} +  \frac{\partial}{\partial x}\int_{-\pi}^{\pi} \alpha^2 \sin(p)\cos(p) \, f(x,p,t) \, dp  + j  = Q(f),
\end{equation}
where
\begin{equation}
Q(f)  = \int_{-\pi}^\pi \frac{ \alpha}{\tau} \cos(p) \int_{-\pi}^\pi\, b(p,p')\left[ M(p) f(p') - M(p') f(p) \right] dp' \, dp.
\end{equation}
Equations \eqref{ME1}, \eqref{ME2} and \eqref{ME3} constitute an Euler-like system of macroscopic equations for the unknowns $\rho$, $j$ and $\kappa$.
It is not closed, since it still depends on several averages of $f$ which, in general, cannot be expressed in terms of  $\rho$, $j$ and $\kappa$.
\par
In order to get a more explicit system, we introduce a further simplification by assuming that the scattering is isotropic, i.e.,
\begin{equation}
  b(p,p') = 1.
\end{equation}
In this case, the collisional term at the right-hand side of the transport equation \eqref{BE} takes the simple relaxation-time form
\begin{equation}
\label{BGK}
   \frac{1}{\tau}\int_{-\pi}^\pi \left[ M(p) f(p') - M(p') f(p) \right] = \frac{1}{\tau}\left[M(p)\rho - f(p) \right],
\end{equation}
also known as BGK (Bhatnagar-Gross-Krook \cite{BGK54}) operator (as usual, the variable $x$ and $t$ are omitted in the description of collisions).
It is readily seen that, with such assumption, system \eqref{ME1}-\eqref{ME2}-\eqref{ME3} assumes the simpler form
\begin{align}
\label{EE1}
&\frac{\partial \rho }{\partial t} + \frac{\partial j}{\partial x} = 0,
\\[4pt]
\label{EE2}
&\frac{\partial j }{\partial t}  +   \frac{\partial}{\partial x} \int_{-\pi}^{\pi} \alpha^2 \sin^2(p)\,f(x,p,t)\,dp - \kappa = 
- \frac{ j}{\tau} 
\\[4pt]
\label{EE3}
&\frac{\partial k}{\partial t} + \frac{\partial}{\partial x}\int_{-\pi}^{\pi} \alpha^2 \sin(p)\cos(p) \, f(x,p,t) \, dp  + j  = \frac{\varepsilon  \rho - \kappa }{\tau},
\end{align}
where $\varepsilon$ is a constant that (in the assumption $\sigma \ll 1$) is given by 
\begin{equation}
\varepsilon =  \int_{-\pi}^{\pi}\alpha \cos(p)\,M(p) \, dp \approx \int_{-\infty}^{+\infty}  \alpha \cos(p)\,M(p) \, dp
= \alpha\,e^{-\sigma^2/2}.
\end{equation}
Of course, this system is still non closed, since the currents in eqs.\ \eqref{EE2} and \eqref{EE3} are extra moments of $f$. 
However we can notice that the system contains (and in the space-homogeneous case reduces to) a damped harmonic oscillator with a forcing term, 
namely
$$
\left\{
\begin{aligned}
&\frac{\partial j }{\partial t}  - \kappa = - \frac{ j}{\tau},
\\[4pt]
&\frac{\partial k}{\partial t} + j = \frac{\varepsilon  \rho - \kappa }{\tau},
\end{aligned}
\right.
$$
which is a manifestation of Bloch oscillations. 
Indeed, such a feature of this peculiar Euler system comes from the fact that the velocity is a sinusoidal function of $p$, which is exactly what 
makes the term $+j$ appear in eq.\ \eqref{rEE3} (coming form the second derivative of the velocity).
We will come back on this point in Sec.\ \ref{S4}, but first let us investigate the behaviour of the system for long times.
\subsection{Diffusion asymptotics}
\label{S3.2}
By looking at the right-hand side of eq.\ \eqref{EE2} one notices that collisions tend to relax the current to zero, which obviously due to the 
fact that the collisions we are describing conserve the mass but not the momentum.
So, it is natural to consider the diffusion asymptotics, which is obtained by looking at the system on a 
time scale which is much larger than the typical collision time \cite{J2009}.
The diffusive scaling is therefore obtained by assuming $\tau \ll 1$ and by rescaling the time as follows:
$$
  t \longmapsto \frac{1}{\tau}\,t.
$$                       
The BGK transport equation and the Euler system are then re-written with such a rescaled time:
\begin{align}
\label{rBE}
&\tau^2 \frac{\partial f}{\partial t} + \tau \alpha \sin (p)  \frac{\partial f}{\partial x}  +  \tau \frac{\partial f}{\partial p} 
=  M(p)\rho - f(p),
\\[4pt]
\label{rEE1}
&\tau \frac{\partial \rho }{\partial t} + \frac{\partial j}{\partial x} = 0,
\\[4pt]
\label{rEE2}
&\tau^2 \frac{\partial j }{\partial t}  +  \tau \frac{\partial}{\partial x} \int_{-\pi}^{\pi} \alpha^2 \sin^2(p)\,f(x,p,t)\,dp - \tau\kappa = - j,
\\[4pt]
\label{rEE3}
&\tau^2\frac{\partial k}{\partial t} + \tau\frac{\partial}{\partial x}\int_{-\pi}^{\pi} \alpha^2 \sin(p)\cos(p) \, f(x,p,t) \, dp  + \tau j  = \varepsilon  \rho - \kappa.
\end{align}
The unknowns are now expanded in powers of the small parameter $\tau$
$$
\begin{aligned}
  &f  = f^{(0)} + \tau f^{(1)} +  \tau^2 f^{(2)} + \cdots, & \quad
  &\rho = \rho^{(0)} + \tau \rho^{(1)} +  \tau^2 \rho^{(2)} + \cdots,
\\
   &j  = j^{(0)} + \tau j^{(1)} +  \tau^2 j^{(2)} + \cdots, & \quad
  &\kappa = \kappa^{(0)} + \tau \kappa^{(1)} +  \tau^2 \kappa^{(2)} + \cdots,
\end{aligned}
$$
and these expansions are inserted in eqs.\ \eqref{rBE}--\eqref{rEE3}. 
At leading order we obtain
\begin{equation}
\label{leading}
   f^{(0)} = M\rho^{(0)}, \qquad  j^{(0)} = 0, \qquad \kappa^{(0)} = \varepsilon  \rho^{(0)}
\end{equation}
which are mutually consistent, since $ f^{(0)}$ is the equilibrium Maxwellian which does not carry current.
At order $\tau$ in \eqref{rEE1}--\eqref{rEE3} we get 
\begin{equation}
\begin{aligned}
&\frac{\partial \rho^{(0)}}{\partial t} + \frac{\partial j^{(1)} }{\partial x} = 0,
\\[4pt]
& j^{(1)} = - \frac{\partial}{\partial x} \int_{-\pi}^{\pi} \alpha^2 \sin^2(p)\,f^{(0)}(x,p,t)\,dp + \kappa^{(0)},
\\[4pt]
& \kappa^{(1)} = - \frac{\partial}{\partial x}\int_{-\pi}^{\pi} \alpha^2 \sin(p)\cos(p) \, f^{(0)}(x,p,t) \, dp  - j^{(0)}  + \varepsilon  \rho^{(1)}.
\end{aligned}
\end{equation}
The first two identities, together with \eqref{leading}, tell us that, up to $\mathcal{O}(\tau)$,  $\rho$ obeys the drift-diffusion equation
\begin{equation}
\label{DDE}
 \frac{\partial \rho}{\partial t} = \frac{\partial}{\partial x} \left( D  \frac{\partial \rho }{\partial x}  -  \varepsilon  \rho \right),
\end{equation}
where the diffusion coefficient is given by
\begin{equation}
 D = \int_{-\pi}^{\pi} \alpha^2 \sin^2(p)\,M(p)\,dp \approx  \int_{-\infty}^{+\infty} \alpha^2 \sin^2(p)\,M(p)\,dp 
 = \alpha^2\,\frac{1-e^{-2\sigma^2}}{2}. 
\end{equation}
The third identity reduces to $\kappa^{(1)} = \varepsilon  \rho^{(1)}$, because $f^{(0)}$ is an even function of $p$ and $j^{(0)} = 0$, and 
will be of no use here.
It is an easy exercise to compute the stationary solution of eq.\ \eqref{DDE}, by assuming that at both sides of the semiconductor the 
electron density has a constant value $\rho_c$ (e.g.\ the electronic density in the metal contacts).
This leads to a classical ohmic law for the stationary diffusion current $J_\mathit{diff}$:
$$
 J_\mathit{diff} = \varepsilon \rho_c,
$$
or, substituting back the dimensional variables, 
\begin{equation}
\label{Jdiff}
 J_\mathit{diff} = \frac{F\tau_0 E_0 \rho_c}{p_0^2}\,e^{-\frac{k_BT}{2E_0}}\,
\end{equation}
(where, of course, $J_\mathit{diff}$ and $\rho_c$ indicate the corresponding dimensional variables). 
\par
Equation \eqref{DDE} indicates that a current can actually flow in our device.
On the other hand, it also  reflects the fact that, on the diffusive time scale, our system shows a completely classical behviour, and every trace of the quantum 
dynamics, represented by the BO, is lost.
Therefore, in order to observe the Bloch oscillations we have now to shift back our attention to the shorter time scale, which will be done in next section.
\section{Averaging over space}
\label{S4}
Let us consider the space averages of the hydrodynamic variables $\rho$, $j$ and $\kappa$ introduced in the last section:
 \begin{equation} 
N(t) := \int_0^1\rho (x,t)\, dx,
\quad
J(t) :=  \int_0^1  j(x,t)\, dx,
\quad
K(t) :=  \int_0^1  \kappa(x,t)\, dx
\end{equation}
(we recall that the non-dimensional space variable $x$ varies in the interval $[0,1]$, while its dimensional counterpart varies in $[0,L]$).
Let us make the reasonable assumption that the total flux of $\rho$ through the device boundaries is zero, so that there is no charge accumulation
nor depletion in the device. 
Then, by integrating both sides of equation \eqref{EE1} with respect to $x \in [0,1]$, we obtain 
\begin{equation}
\label{IEE1}
\frac{dN}{dt} = 0,
\end{equation} 
which obviously means that the total number of electrons (and, therefore, the total charge) is conserved.
By also assuming that the total fluxes of  $j$ and $\kappa$ through the boundaries is zero, the integration of \eqref{EE2}  and \eqref{EE3},
with respect to $x \in [0,1]$ yields the system of ODEs
\begin{equation}
\label{IEE23}
\begin{aligned}
&\frac{d J}{dt}   -K = -\frac{1}{\tau}\,J \,
\\[4pt]
&\frac{dK}{d t} + J  = - \frac{1}{\tau}K +  \frac{\varepsilon  N_0}{\tau} \, ,
\end{aligned}
\end{equation}
where $N_0$ is the total number of particles, which is constant, according to eq.\ \eqref{IEE1}.
Hence we see that the averaged hydrodynamic variables $J(t)$ and $K(t)$ behave as a damped harmonic oscillator with a forcing term.
System \eqref{IEE23} can be readily recast into the following second-order equation for $J$
\begin{equation}
\label{EDO}
\frac{d^2 J}{dt^2}+\frac{2}{\tau}\frac{d J}{dt} +\left(1+\frac{1}{\tau^2}\right) J = \frac{\varepsilon N_0}{\tau}.
\end{equation}
Solving this equation is a standard exercise in ODEs, and the solution is
\begin{equation}
\label{J}
J(t) = e^{-t/\tau}[a \cos(t) +b\sin (t)] + \frac{\tau \varepsilon N_0}{ 1+ \tau^2},
\end{equation}
with $a$ and $b$ to be determined from the initial conditions $J_0$ and $K_0$.
By writing this formula in physical variables and substituting $a$ and $b$ with their expressions in terms of the initial 
conditions we obtain
\begin{equation}
\label{Jphys}
J(t) = e^{-t/\tau_0}\left[\left(J_0 - J_\infty\right ) \cos(\omega t) + \left(K_0 - \frac{1}{\omega\tau_0}J_\infty \right)\sin (\omega t)\right] + J_\infty,
\end{equation}
where 
\begin{equation}
   \omega = \frac{1}{t_0} = \frac{F}{p_0}
\end{equation}
is the BO frequency, 
\begin{equation}
\label{Jinfty}
 J_\infty = \frac{F\tau_0 E_0N_0}{p_0^2 + (F\tau_0)^2}\,\,e^{-\frac{k_BT}{2E_0}}
\end{equation}
is the asymptotic value of the current and
\begin{equation}
\begin{aligned}
&J_0 = \int_0^L\int_{-\pi p_0}^{\pi p_0} \frac{E_0}{p_0}\sin\left(\frac{p}{p_0}\right)  f_{\mathrm{in}}(x,p) \, dp \, dx,  
\\[4pt]
&K_0 = \int_0^L \int_{-\pi p_0}^{\pi p_0}  \frac{E_0}{p_0} \cos\left(\frac{p}{p_0}\right) f_{\mathrm{in}}(x,p) \, dp \, dx. 
\end{aligned}
\end{equation}
are the initial values of $J$ and $K$ (in physical variables) obtained from the initial phase-space distribution $ f_{\mathrm{in}}$.
We see from \eqref{Jphys} that, after a transient which is dominated by Bloch oscillations, the current reaches the stationary value $J_\infty$.
\par
It is worth noticing here that the difference between $J_\infty$ and the stationary diffusive current $J_\mathit{diff}$ (see \eqref{Jdiff}) is due to the fact 
that the diffusive limit is {\em not} a limit for large times, but its is a time-scale asymptotic for $\tau \ll 1$ (that is $\tau_0 \ll p_0/F$), 
which is not necessarily the case in general.

%

We can notice that in both limits $\tau \to \infty$ and $\tau \to 0$ the current vanishes. 
The behaviour when $\tau$ approaches  0 (which means exactly that there are many collisions in the time it takes $p$ to span the band period) 
is close to the classical ohmic regime (and indeed $J_\infty \approx J_\mathit{diff}$ becomes a linear function of $F$).
So, the limit  $\tau \to 0$ corresponds to the resistivity becoming infinite.
The behaviour for large $\tau$, instead, corresponds to the opposite regime in which the collisions are rarefied, compared to the band-spanning time,  
and electrons tend to remain trapped in the periodic trajectories. 
The limit $\tau \to \infty$ is therefore a clear illustration of the Bloch paradox: without collisions no current can flow.
\section{Comparison with experiments and discussion}
\label{S5}
In order to discuss the possibility of comparing the formula \eqref{Jphys} with experimental measurements, we first have to make some considerations
about the band parameter $p_0$.
Assume that the periodic structure has a period $d$. 
Then the period of the reciprocal lattice (i.e.\ the wavenumber) is $k_0 = 2\pi/d$. 
Now, the De Broglie identity implies that the corresponding momentum period ($2\pi p_0$) is given by $ 2\pi p_0 = \hbar k_0$,
where $\hbar = h/2\pi$ and $h$ is the Planck constant.
Then, we have a simple relation between the lattice period and $p_0$:
\begin{equation}
\label{p0_vs_d}
  p_0 = \frac{\hbar}{d} .
\end{equation}
As a consequence, the BO frequency as a function of the applied voltage is given by
\begin{equation}
  \frac{\omega}{2\pi} =  \frac{F}{2\pi p_0} = \frac{qV d}{h L},
\end{equation}
and the non-dimensional parameter $\tau$, that is the ratio between the collision time and the band-spanning time, is given by
\begin{equation}
   \tau =  \frac{\tau_0 F}{p_0}  =  \frac{ \tau_0q V d}{L\hbar}.
\end{equation}
In natural crystals this number is very small, which means that Bloch oscillations are extremely difficult to observe.
Indeed, all experimental results are aimed at observing Bloch oscillations in semiconductor superlattices (SL), which are artificial periodic structures
made by repetitions of several layers of different semiconductors \cite{DOK1995,Leo92}.
Here, the periodicity $d$ is several tenths of the bulk crystal periodicity, resulting in a smaller $p_0$ and a larger $\tau$.
Unfortunately, our simple model is inadequate to describe such a device, since its modelisation would involve different time and space scales and more complicated 
collisional interactions.
\par
\begin{figure}
\begin{center}
\includegraphics[scale=.85]{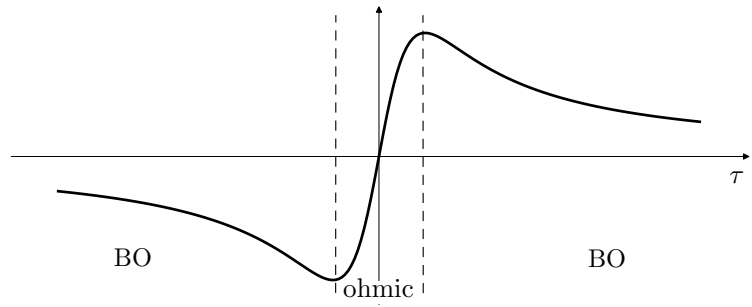}
\caption{A representation of the asymptotic current $J_\infty$ (arbitrary units) as a function of the non-dimensional parameter $\tau$. 
In the central region, $-1 < \tau < 1$, the current exhibits a ohmic behaviour (recall that $\tau$ is proportional to $V$) but the BO cannot be observed because
the damping time is less than the band period. 
For $\vert\tau\vert > 1$ the BO can be observed, but the behaviour is non-ohmic since the current decreases for increasing $V$ (the so-called ``negative differential resistance'').}
\label{fig2}
\end{center}
\end{figure}

The major weakness of our model is that the Ohmic behaviour ($\tau < 1$) and a damping time large enough for the BO to be observed ($\tau > 1$) are 
mutually excluded (see Figure \ref{fig2}).
In the SL experiments, actually, the two conditions are met at the same time. 
This limitation of our model can be fixed ``by hands'' with the substitution of the asymptotic current $J_\infty$ \eqref{Jinfty} 
with the diffusive current $J_\mathit{diff}$ \eqref{Jdiff}.
This modification leads to a surprisingly good agreement the with the experimental results, as it is shown below.
\par
\begin{figure}
\begin{center}
\includegraphics[scale=.65]{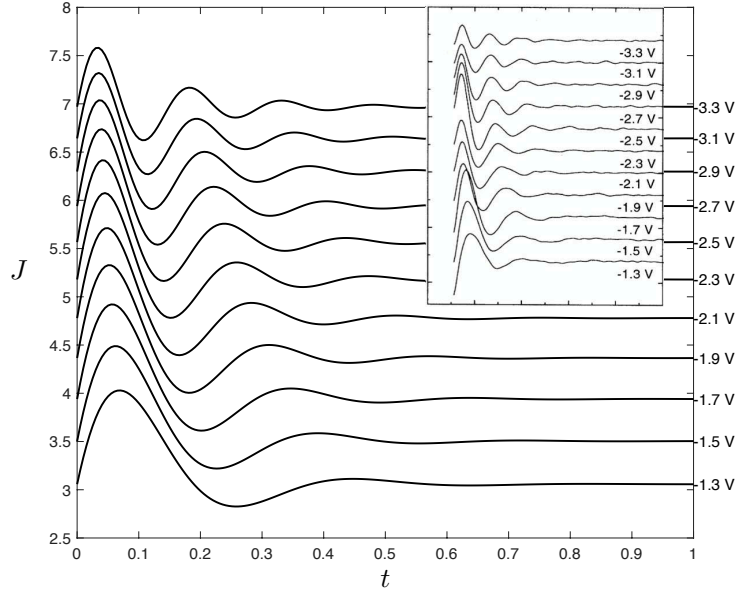}
\caption{Plots of the current as a function of time for several values of the applied voltage. 
The time unit is $1\,\mathrm{ps}$, while the current is in arbitrary units (because it cannot be directly compared with the experimental values, 
which are obtained via polarization measures). 
The temperature is $300\,\mathrm{K}$.
In the inset the corresponding experimental measurements are shown (figure reproduced from Ref.\ \cite{DOK1995} with permission) and
the time unit is the same (time goes from $0$ to $1\,\mathrm{ps}$).}
\label{fig3}
\end{center}
\end{figure}
\begin{figure}
\begin{center}
\includegraphics[scale=.65]{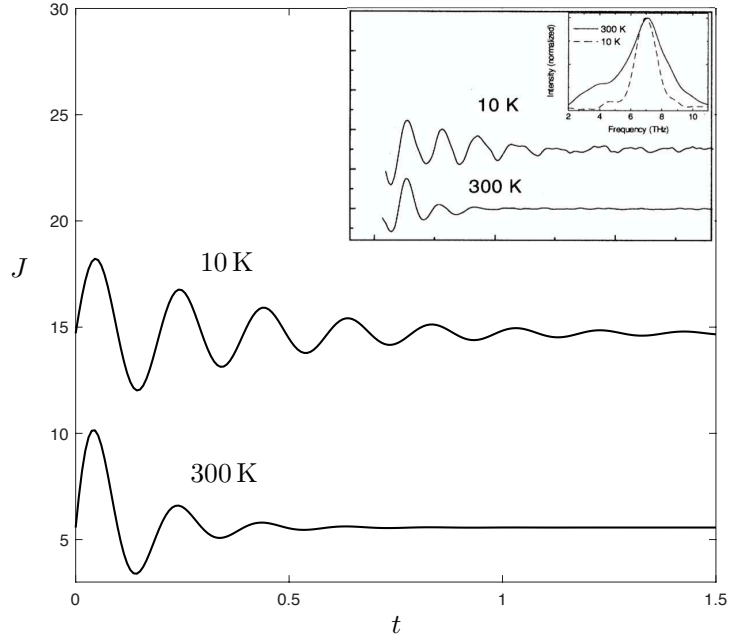}
\caption{Plots of the current as a function of time for two different values of the temperature.  
The applied voltage is $2.5\,\mathrm{V}$ and  the axis units are as in Fig.\ \ref{fig3}.
In the inset the corresponding experimental measurements are shown (reproduced from Ref.\ \cite{DOK1995} with permission) and
the time unit is the same (time goes from $0$ to $1.5\,\mathrm{ps}$). 
The experimental figure contains another inset with the distribution in frequency of
the observed BO.}
\label{fig4}
\end{center}
\end{figure}

According to the experimental device described in Ref.\ \cite{DOK1995}, we choose the following values of the physical parameters:
\begin{itemize}
\item device length: $L =  1.0\times10^{-6}\,\mathrm{m}$;
\item lattice period: $d = 8.4\times 10 ^{-9}\,\mathrm{m}$;
\item band width: $E_0 = 3.6\times 10 ^{-2}\,\mathrm{eV}$;
\item mean collision time at $T = 10\,\mathrm{K}$ : $\tau_0 = 3.7\times 10 ^{-13}\,\mathrm{sec}$;
\item mean collision time at $T = 300\,\mathrm{K}$ : $\tau_0 = 1.3\times 10 ^{-13}\,\mathrm{sec}$;
\item cross-sectional electronic density: $n = 2.0\times 10 ^{9}\,\mathrm{cm^{-2}}$.
\end{itemize}
The value $p_0 = 7.86 3.7\times 10 ^{-13}\,\mathrm{sec}$ is calculated by means of \eqref{p0_vs_d}. 
These values are used in formula \eqref{Jphys} (with $J_\mathit{diff}$ in place of $J_\infty$) to compute $J(t)$ for different values of voltage and temperature.
In Figure \ref{fig3}, $J(t)$ is plotted for several values of the applied potential $V$ and compared with the analogous experimental figure, shown in the inset.
In Figure \ref{fig4}, $J(t)$ is plotted for two different values of the temperature and again compared with the corresponding experimental figure.
In both case we can observe a good qualitative agreement of our (modified) model with the real data.
\par
In conclusion, we have proposed a reasonably simple mathematical model with the aim of illustrating the ``paradox'' of Bloch oscillations. 
In spite of its simplicity, the model is able to give qualitatively good results when comparing its predictions with experimental data.
Of course, in order to have an accurate description of a real device, a more refined model is needed, especially with regards to the description 
of collisions.
There are many available kinetic model of electron-phonon interaction (see e.g.\ \cite{CMR2020}) that could be used to improve the transport equation 
\eqref{Boltzmann} and  possibly reproduce the experimental results without any {\em ad hoc} assumptions.

\section*{Acnowledgements}
The author wishes to thank his team of students at the XI {\em Modelling Week} of Universidad Complutense de Madrid, 
Antonio Luna, Carlos Domingo, Jaime Oliver, Jared M. Field, Rahil Sachak-Patwa and Sergio Montoro, 
where the first nucleus of this work originated: .


\begin{thebibliography}{99}

\bibitem{AM1976}
N.W. Ashcroft, N.D.Mermin.
\textit{Solid State Physics}.
Saunders College Publishing, 1976.

\bibitem{LB2000}
Barletti, L., 2000.
Some remarks on affine evolution equations with applications to particle transport theory.
{\it Math. Meth. Models Appl. Sci.} 10:877--893.

\bibitem{CMP11}
L. Barletti, N. Ben Abdallah. 
Quantum transport in crystals: effective-mass theorem and k$\cdot$p Hamiltonians.
{\em Comm. Math. Phys.} {\bf 307}, 567--607  (2011).

\bibitem{BGK54}
Bhatnagar, P. L., Gross, E. P., Krook, M. (1954). A model for collision processes in gases.
I. Small amplitude processes in charged and neutral one-component systems. {\it Phys.
Rev.} 94:511?525.

\bibitem{Belleni79}
Belleni-Morante A.
\newblock \textit{Applied Semigroups and Evolution Equations}.
\newblock Clarendon Press: Oxford, 1979.

\bibitem{BJ2000}
B. Bransden and C. Joachain. {\em Quantum Mechanics}.
Prentice-Hall, 2000.

\bibitem{CMR2020} 
\newblock V. D. Camiola, G. Mascali and V. Romano, 
\newblock {\it Charge Transport in Low Dimensional Semiconductor Structures}, 
\newblock Springer,  2020.

\bibitem{DOK1995}
T. Dekorsy, R. Ott, and H. Kurz. {\em Bloch oscillations at room temperature.} Physical Review B 51 (1995), 23, pp. 17275-17278. 

\bibitem{Frensley87}
W. R. Frensley. Wigner-Function Model of a Resonant-Tunneling Semiconductor
Device. {\it Physical Review B} 36.3 (1987), pp. 1570?1580.

\bibitem{J2009}
J\"ungel, A.: Transport Equations for Semiconductors. Springer, Berlin (2009)

\bibitem{Leo92}
Leo, K., Bolivar, P. H., Br\"uggemann, F.; Schwedler, R., K\"ohler, K. (1992). 
Observation of Bloch oscillations in a semiconductor superlattice. 
{\em Solid State Communications.} 84 (10), 943--946.

\bibitem{TilliTesi}
D. Tilli, Master thesis, Universit\`a di Firenze (2011)


\end{thebibliography}
\end{document}